\documentclass[10pt,aps,prl,twocolumn,showpacs,superscriptaddress,nofootinbib]
{revtex4-1}

\usepackage{amssymb}
\usepackage{graphicx}

\begin{document}

\title{Mechanical Rainbow Trapping and Bloch Oscillations in Structured Elastic Beams}

\author{A. Arreola-Lucas}
\affiliation{Posgrado en Ciencias e Ingenier\'ia de Materiales, Universidad Aut\'onoma Metropolitana-Azcapotzalco, 02200 M\'exico Distrito Federal, Mexico}
\affiliation{Wave Phenomena Group, Universitat Polit\`ecnica de Val\`encia, Camino de vera s.n. (Building 7F), ES46022, Valencia, Spain}
\author{G. B\'aez}
\affiliation{Departamento de Ciencias B\'asicas, Universidad Aut\'onoma Metropolitana-Azcapotzalco, 02200 M\'exico Distrito Federal, Mexico}

\author{F. Cervera} 
\author{A. Climente} 
\affiliation{Wave Phenomena Group, Universitat Polit\`ecnica de Val\`encia, Camino de vera s.n. (Building 7F), ES46022, Valencia, Spain}

\author{R. A. M\'endez-S\'anchez}
\affiliation{Instituto de Ciencias F\'isicas, Universidad Nacional Aut\'onoma de M\'exico, Apartado Postal 48-3, 62210 Cuernavaca Mor., Mexico}

\author{J. S\'anchez-Dehesa}
\email{jsdehesa@upv.es}  
\affiliation{Wave Phenomena Group, Universitat Polit\`ecnica de Val\`encia, Camino de vera s.n. (Building 7F), ES46022, Valencia, Spain}

\begin{abstract}

We demonstrate, both experimentally and numerically, the mechanical rainbow trapping effect and the mechanical Bloch oscillations for torsional waves propagating in chirped mechanical beams.
After extensive simulations, three quasi-one-dimensional chirped structures were designed, constructed and experimentally characterized by Doppler spectroscopy. 
When the chirp intensity vanishes a perfect periodic system, with bands and gaps, is obtained. 
The mechanical rainbow trapping effect occurs for small values of the chirp intensity. 
The wave packet traveling along the beam is progressively slowing down and is reflected back at a certain depth, which depends on its central frequency. 
In this case a new kind of oscillation, here named ``{\em rectified rainbow-Bloch oscillation}'', appears since the wave packet is reflected at one side by the interface between the structure and the uniform rod and by the minigap at the opposite side.
For larger values of the chirp parameter the rainbow trapping yields the penetration length where the mechanical Bloch oscillations emerge.
Numerical simulations based on the transfer matrix method are in agreement with experimental data.
\end{abstract}

\date{\today}

\maketitle

The control of electromagnetic (EM) waves by artificial structures named EM metamaterials has been accompanied by the development of acoustic and mechanical metamaterials, with the goal of controlling sound and vibrations, respectively. 
Excellent review articles~\cite{KadicRPP2013,CummerNat2016} are already available reporting the amazing properties of these new type of manmade structures.
Though many exciting phenomena, as the rainbow trapping effect and the analogue of the electronic Bloch oscillations, have been described and experimental demonstrated for acoustic metamaterials, the experimental demonstration for elastic systems is still scarce. 
Thus, novel structures for the control of elastic waves are mainly based on theoretical proposals since vibrations usually involve a mixture of polarizations, which implies difficulties of measuring them selectively.

The rainbow trapping effect, on the one hand, is one of the most interesting phenomenon recently discovered in the field of optics~\cite{Tsakmakidis}.
In this effect the wave packets are slowed down up to different spatial depths, within a synthetic structure that has embedded a metamaterial with negative refractive index. 
The reached spatial depths depend on the central frequency of the wave packet. 
Since its discovery in 2007, many potential applications have been reported~\cite{Gan1,Park,Zhao,Smolyaninova,Hu,Shen1,Shen2,Gan2,Khurgin}. 
Thus, the light can be trapped or multiplexed using this effect~\cite{Sun,Jang,Chen2,Bouillard,Cui}. 
A demonstration of its acoustic analogue has been reported by using acoustic metamaterials~\cite{Zhu2013,Ni} and phononic crystals~\cite{RomeroGarciaEtAl}. 
More recently, the trapping of Lamb waves has been theoretically reported~\cite{tian}. 
On the other hand, electronic Bloch oscillations was predicted in semiconductors when a dc electric field is applied~\cite{Ascrhoft}.  
This quantum phenomenon was demonstrated in the late eighties thanks to the discovering of semiconductor superlatices~\cite{MendezEtAl}. 
Later, the analogue of Bloch oscillations has been also shown in different structures supporting the propagation of other types of waves, like dielectric structures~\cite{SapienzaEtAl,AgarwalEtAl} and waveguides~\cite{MorandottiEtAl}, ultracold atoms~\cite{BattestiEtAl}, phononic crystals~\cite{Sanchis-AlepuzEtAl,HeEtAl,deLimaEtAl}, superconducting nanowires~\cite{LehtinenEtAl} and more recently in molecular motion~\cite{FlossEtAl}. 
Though Bloch oscillations are expected to occur on vibrating structures, up to now, there is no demonstration of this effect for mechanical waves. 

In this Letter we demonstrate the emergence of both phenomena, Bloch oscillations and rainbow trapping, in elasticity. This is done for torsional waves propagating in a mechanical structure consisting of a metallic beam with notches. 
Bloch oscillations are obtained by mimicking the dc electric field using a gradient of the cavity thicknesses determined by the notches.  
The gradient defines the so called chirping parameter of the resulting structure. 
For non-vanishing values of the chirping parameter the group velocity of the mechanical waves depends on the frequency and the position inside the beam. 
Thus, wave packets with different frequencies are back-scattered at different positions inside the corrugated beam; being packets with higher frequencies the ones with larger penetration depths. 
This behavior represents the mechanical analogue of the rainbow trapping effect earlier demonstrated for optical and acoustical waves. 
Since part of the back-scattered wave packet is returned by the mechanical analog of the electric field, the Bloch oscillations emerge. The wave packet then oscillates with a period given by the inverse of the width of the passing band.
\begin{figure}
\includegraphics[width=1.0\columnwidth]{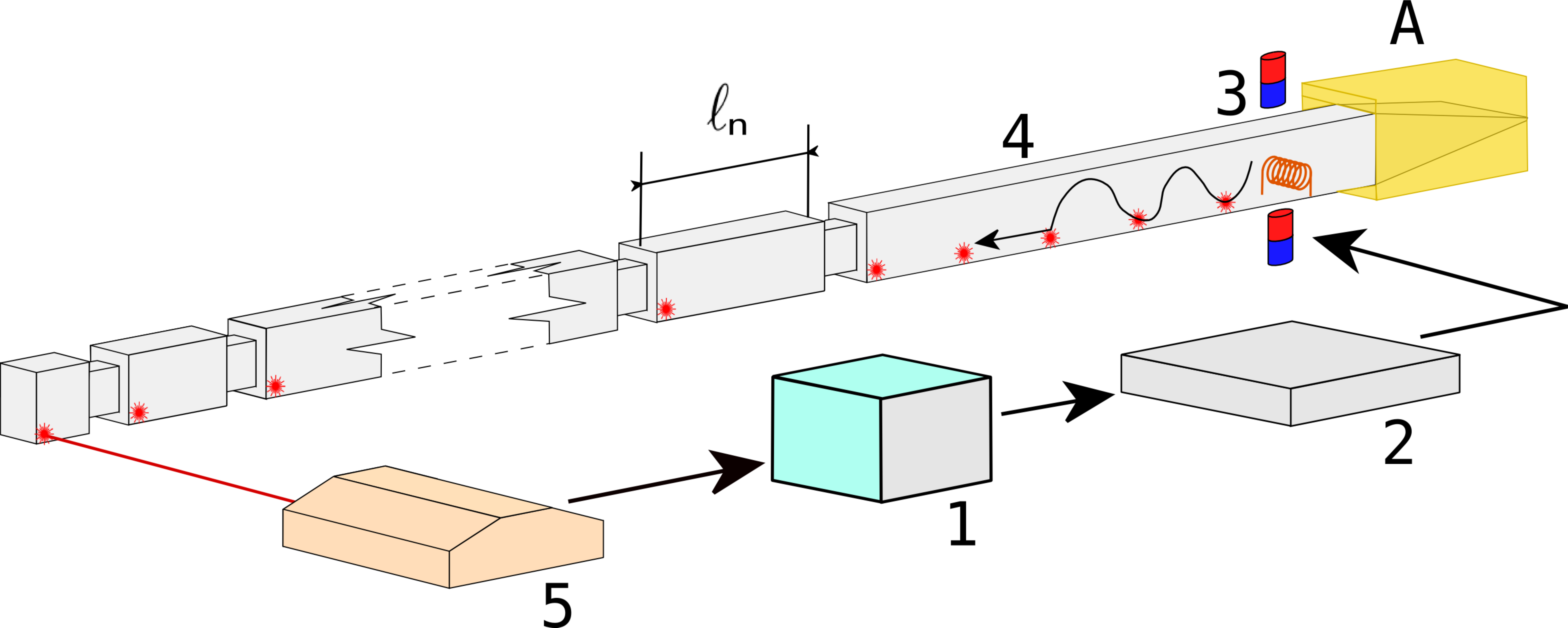}
\caption{(Color online) Structured metallic beam and experimental setup: (1) NI-PXI for generating, recording and analyzing signals, (2) high-fidelity audio amplifier, (3) electromagnetic-acoustic transducer, (4) machined beam and (5) Doppler interferometer. 
The beam with rectangular section can be separated in three regions. 
On the right, a vibration isolation system (A) consisting of a wedge covered by an absorbing mastic seal. 
The central region is uniform while the part on the left contains the chirped structure; the lengths $\ell_n$ are determined with Eq.~(\ref{Eq.ell}). 
The (red) spots indicate the position of the laser measurements.}
\label{fig1}
\end{figure}

As illustrated in Fig.~\ref{fig1}, the beam can be considered as divided in three regions. The structured region is located at one end. 
The central region is uniform while the opposite end contains a passive vibration isolation system (A). 
The chirped system consists of an aluminum beam with rectangular cross section, with height 30~mm and width 10~mm, where the structure is machined; it is composed of 20 cells with varying lengths $\ell_n$ separated with notches using the rule that yields the elastic Wannier-Stark ladders~\cite{GutierrezEtAl,MonsivaisEtAl,GhulinyanEtAl}
\begin{equation}
 \ell_n=\frac{\ell_0}{1+n \gamma}
 \label{Eq.ell}
\end{equation}
for $n=1,\ldots,20$. Here $\gamma$ is the chirp parameter that mimics the electric field, and $\ell_0$ is an arbitrarily defined length determining the actual size of the series.
When $\gamma =$ 0 a periodic structure is obtained but for $\gamma > 0$ ($\gamma < 0$) the structure becomes chirped and cells become smaller (longer) with index $n$~\cite{GutierrezEtAl}. 
\begin{figure}
\includegraphics[width=0.8\columnwidth]{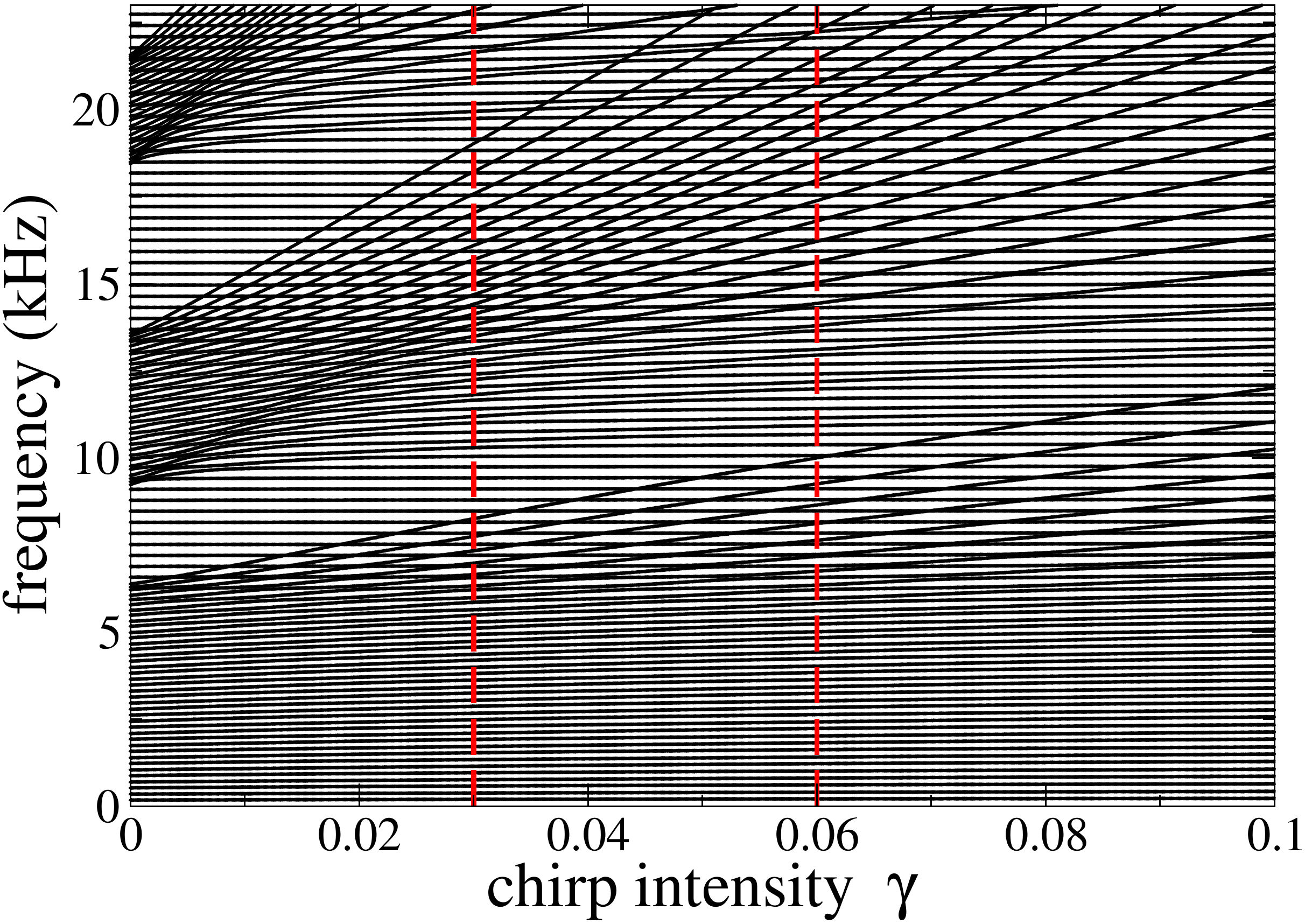}
\caption{(Color online) Band structure of the torsional modes propagating in a structured metallic beam calculated with the transfer matrix method. 
The levels are given as function of the chirp parameter $\gamma$. 
The three structures manufactured and characterized correspond to $ \gamma= 0,\, 0.03$ and $0.06$; the last two are defined by the vertical (red) dashed lines.
}
\label{fig2}
\end{figure}

The frequency spectrum of the torsional modes for a finite-length chirped structure was obtained numerically using the transfer matrix (TM) method~\cite{Morales2}. 
Calculations were performed by considering an aluminum beam with free boundary ends and composed by a uniform part with length 203 cm. 
The chirped part was determined using $\ell_0=$ 9.2cm. 
The level dynamics as a function of the chirp parameter $\gamma$ is shown in Fig.~\ref{fig2}, where it is observed that the frequency spectrum also contains levels having a flat dispersion relation with $\gamma$. 
The flat levels correspond to modes associated to the uniform part of the finite-length beam. 
For $\gamma =0$ the typical frequency spectrum of finite-length beam containing a finite periodic structure is observed. 
Since a free surface is used at both ends of the beam a band starting at zero frequency results. 
From the calculation, the gaps and bands of the perfect periodic structure can be established.  
Thus, for no chirp the first gap is approximately located between 6.5 kHz and 9.5 kHz. 
The latter defines the beginning of the second band, which ends at $\approx 13.5$~kHz. 

For increasing values of $\gamma$, Fig.~\ref{fig2} shows that the discrete levels in a given band start to separate, the bands become wider and the gaps narrower. 
In the second band, three different regimes are clearly defined. 
The first one corresponds to values $\gamma \approx 0$. 
In this regime the level density is inhomogeneous and has maxima close to the borders of the band; this is a reminiscence of the perfect periodic system. 
The second regime appears for $0.03 < \gamma < 0.065$ where the level density inside the band, for a fixed value of $\gamma$, is approximately homogeneous, {\em i.e.}, the levels are almost equally spaced. 
The Wannier-Stark ladders spectrum dominates this regime~\cite{GutierrezEtAl}. 
The third regime appears when nearest neighbors bands start to overlap. 
Thus, after a critical value of the chirp parameter Zenner tunneling between bands should be observed in the form of an enhanced transmission peak \cite{Sanchis-AlepuzEtAl,GhulinyanEtAl}.
Results presented here correspond to the first two regimes.

The time evolution of the wave packets is also studied in the framework of the TM formalism. 
An initial wave packet, located at the free boundary of the uniform part of the beam, is expanded in around 250 normal-mode wave amplitudes and then its evolution is obtained in terms of the stationary solutions.
The numerical algorithm was tested in several cases of interest including the doorway state mechanism in the time domain~\cite{MoralesEtAl2013}.
 
The level dynamics described above together with a comprehensive numerical study of the evolution of wave packets for different $\gamma$ values was employed to select the three structured beams under study here; i.e., $\gamma=0,\, 0.03$, and $0.06$, respectively.
The manufactured sample with $\gamma=$0 contains a periodic region consisting of 20 equal cells with lengths of 92 mm joined by rectangular cuboids (the notches) with height 18.0~mm, width 6.0~mm and length 8.0~mm. 
The samples with chirp parameters $\gamma=0.03$ and $0.06$, respectively, consist of 20 cells with variable length $\ell_n$, determined by Eq.~(\ref{Eq.ell}) with $\ell_0=92$~mm. 
These chirped structures exhibit the two non-intuitive behaviors of interest here; the mechanical rainbow trapping and the mechanical Bloch oscillations, respectively. 
\begin{figure}
\includegraphics[width=1.0\columnwidth]{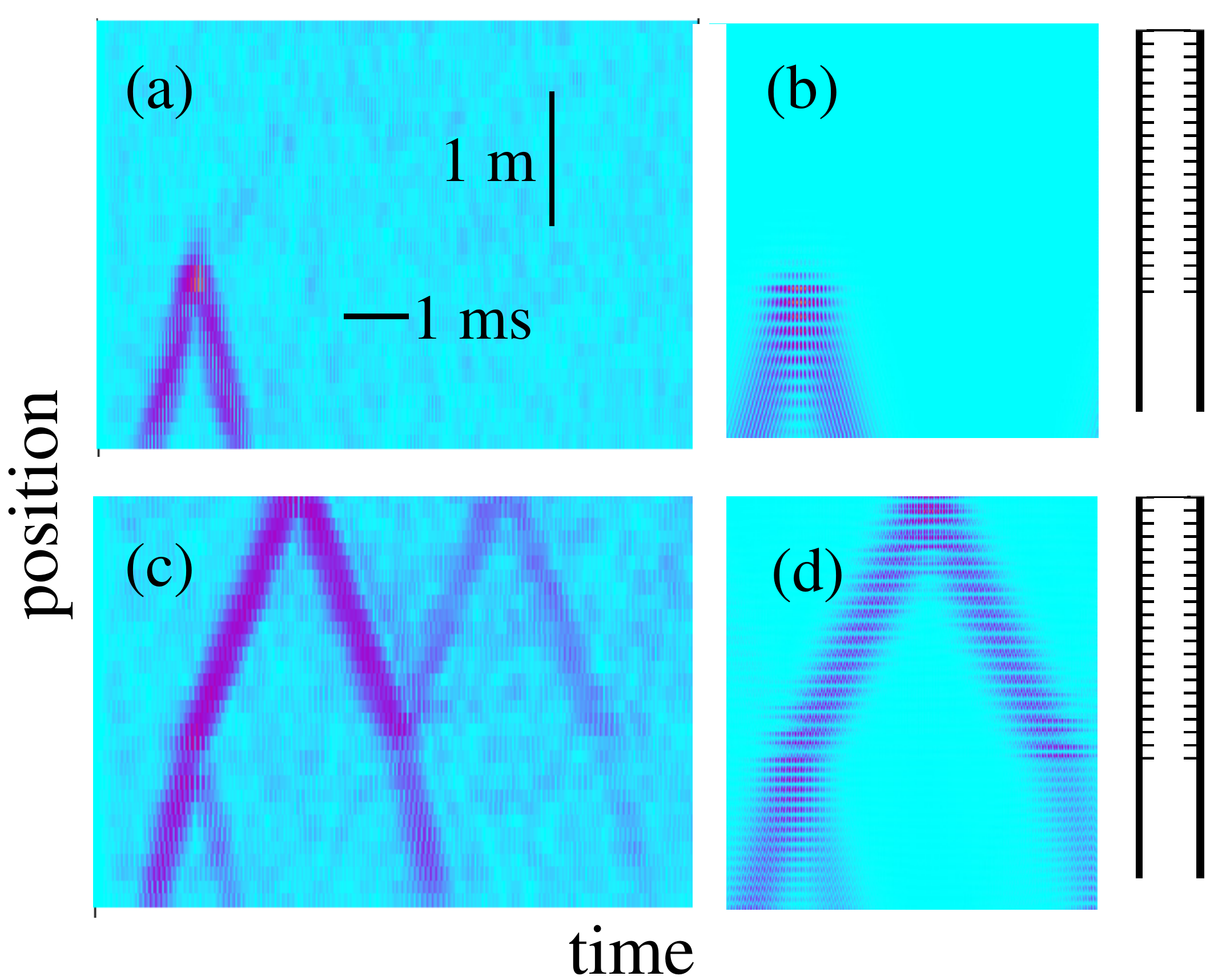}
\caption{(Color online) Wave packet intensity as a function of the position and time for the periodic structure, $\gamma =0$. The left and right columns show the experimental and numerical results, respectively. In cases (a) and (b) the central frequency lies in the first gap ($f_\mathrm{C} = 8$~kHz); in cases (c) and (d) the central frequency lies in the second band ($f_\mathrm{C} = 11.5$~kHz). 
The plots on the right describe the structured beam, showing both the homogeneous and the periodically structured parts.
}
\label{fig3}
\end{figure}

The experimental characterization has been performed as follows. 
First, torsional waves are generated in the uniform part of the beam and sent to the chirped structure using the setup schematically depicted in Fig.~\ref{fig1}.  
The signal generated by a NI-PXI is amplified by a high-fidelity audio amplifier Cerwin-Vega CV5000 and then sent to an electromagnetic-acoustic transducer (EMAT). 
The EMAT generates a torsional Gaussian wave packet at its position, located in the uniform part of the beam. 
The wave packet travels towards the chirped structure and the vibrations are measured with a laser Doppler vibrometer and analized with the PXI. 
In order to reconstruct the dynamics of the wave packets, the measurements are done in several positions along the mechanical structure and in the uniform part of the beam.
For frequencies higher than $\approx 1.5$~kHz, the waves arriving to the opposite side of the beam are almost completely absorbed by a passive vibration isolation system; the latter consists of a wedge covered by an absorbing mastic seal.
Figure~\ref{fig3} shows the dynamics of a Gaussian wave packet propagating in the periodic structure; {\em i.e.} $\gamma=0$ in Eq.~(\ref{Eq.ell}). 
The wave packet with an initial width of $0.5$~ms has a central frequency, $f_\mathrm{C}$, located at two different frequencies: within the gap and inside the band, respectively. 
The width used in the time domain implies than the wave packet has a spatial width of $0.875$~m in the uniform part of the beam since the velocity of the torsional waves in the beam is $c=1750$~m/s.
\par
Figures~\ref{fig3}(a) and~\ref{fig3}(b) show the behavior of the wave packet when its central frequency $f_\mathrm{C}$ lies within the first gap. 
It is observed how the wave packet is completely reflected at the interface between the uniform part and periodic structure of the beam. 
However, when $f_\mathrm{C}$ lies within the second band, Figs.~\ref{fig3}(c) and (d) show that the wave packet is partially transmitted and partially reflected each time that the packet crosses the interface. A good agreement is observed between experimental data (left panels) and numerical simulations (right panels).
\begin{figure}
\includegraphics[width=1.0\columnwidth]{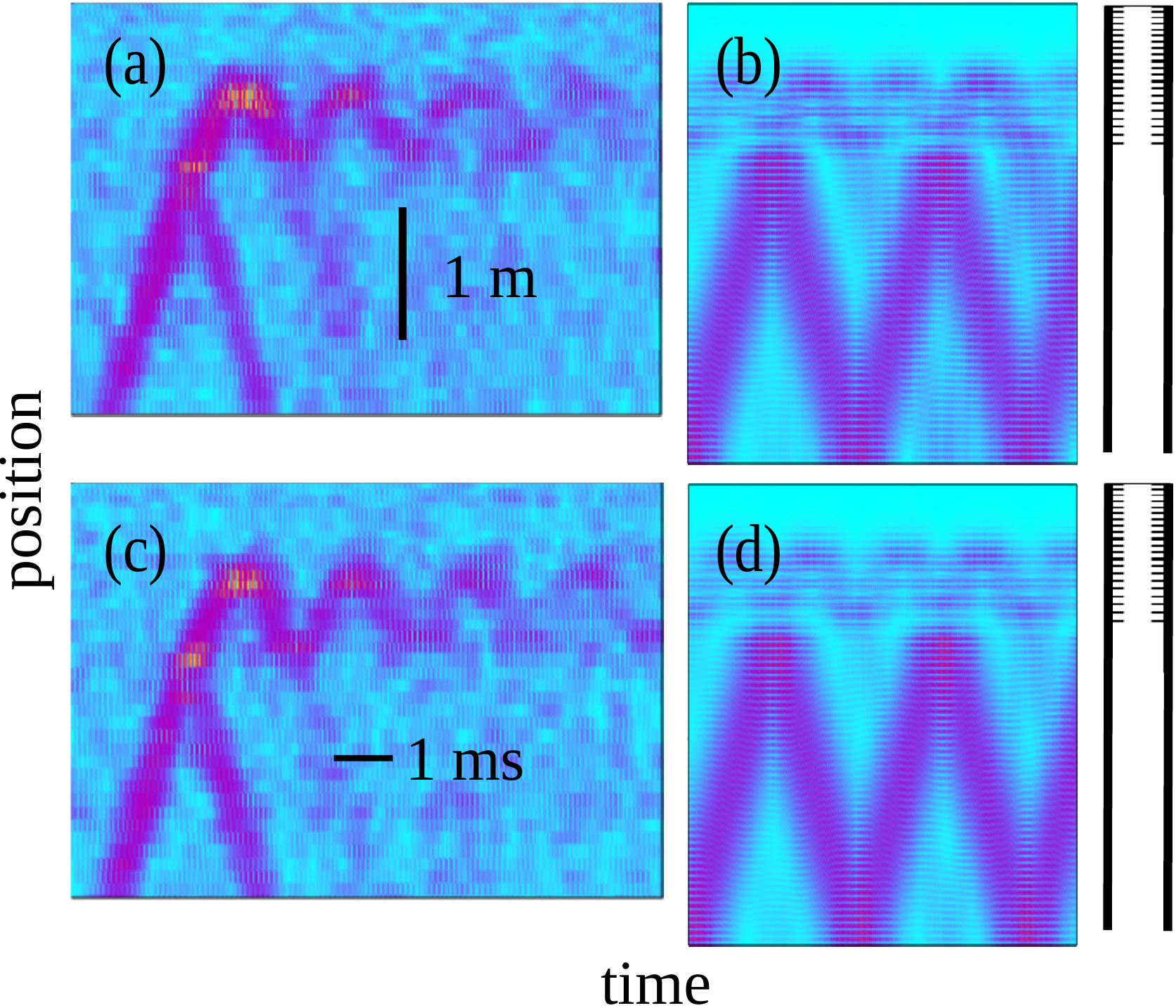}
\caption{(Color online) The Bloch oscillations are apparent in a beam with a chirped structure ($\gamma = 0.06$).
The left and right columns correspond to the experimental and numerical results, respectively. 
In cases (a) and (b) the wave packet has a central frequency $ f_\mathrm{C} = 14.5$~kHz; in cases  (c) and (d) $f_\mathrm{C} = 15.0$~kHz.
 The plots on the right describe the structured beam, showing both the homogeneous and the chirped parts.
}
\label{fig4}
\end{figure}
\begin{figure}
\includegraphics[width=1.0\columnwidth]{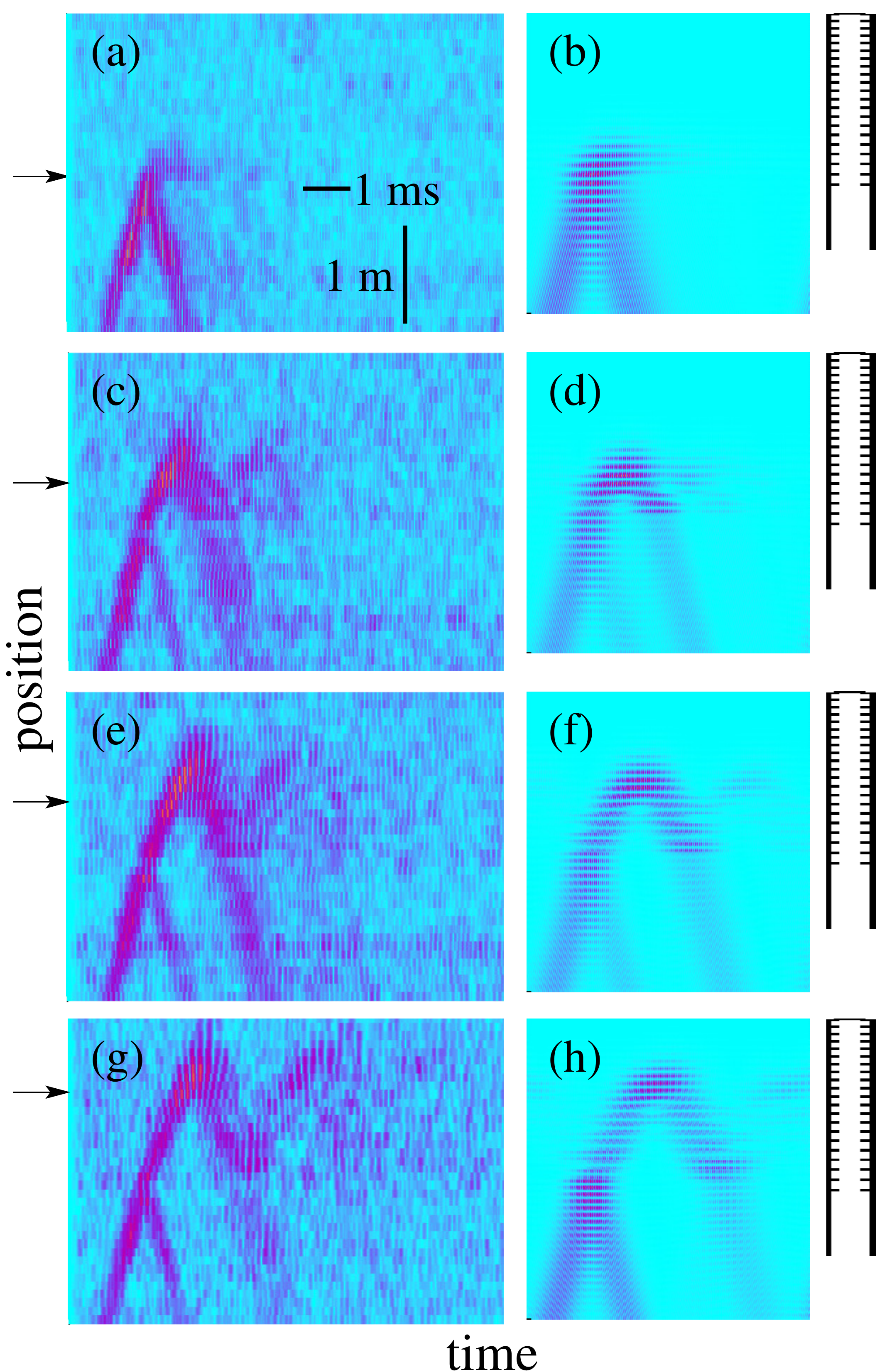}
\caption{(Color online) The rainbow trapping is apparent in a chirped mechanical structure with $\gamma = 0.03$.
The left and right columns correspond to the experimental and numerical results, respectively.
The central frequency of the wave packet is $9$~kHz for cases (a) and (b), $10$~kHz for cases (c) and (d), $11$~kHz for cases (e) and (f), and  $12$~kHz for cases (g) and (h).
The arrows indicate the penetration depth predicted by the independent rod model.}
\label{fig5}
\end{figure}

Let us study now the case of chirped structures where the mechanical analogue of the electronic Bloch oscillations has been characterized. 
Mechanical Bloch oscillations are clearly appearing in the second regime (i.e., for 0.03 $<\gamma<$ 0.065), when the separation between levels in the second band is larger than in the first regime. 
For both regimes ($\gamma >$ 0), a simple model indicates that the spacing between levels is given by:  
\begin{equation}
\label{eq2}
\delta f_\mathrm{B}=\gamma\frac{c}{2\ell_0}.
\end{equation}
For the chirp parameter considered (6$\%$) the predicted separation is $\delta f_\mathrm{B}\approx 610$~Hz, in good agreement with $\Delta f^\mathrm{(TM)}=600$~Hz, which is the frequency separation obtained from the spectrum shown in Fig.~\ref{fig2}. 
The dynamics of the wave packet is shown in Fig.~\ref{fig4}, where an oscillating behavior is observed within the structured part of the beam.
Plots on the left (right) column correspond to the experimental data (numerical simulations) obtained for the propagation of wave packets with two different central frequencies, within the second allowed band.
The measured period of the Bloch oscillation is $\approx 1.6$~ms, in good agreement with the predictions of the TM method $T_\mathrm{B}=1/\delta f_\mathrm{B}^\mathrm{(TM)}\approx 1.66$~ms and the analytical model, for which $T_\mathrm{B}=1/\delta f_\mathrm{B}\approx 1.64$~ms.
This phenomenon is the mechanical version of the electronic Bloch oscillations. 
It is also observed in Fig.~\ref{fig4} that a small portion of the energy of the wave packet leaks towards the uniform part of the beam.
However, the phenomenon is very robust and four oscillations are clearly observed.  
The Bloch oscillations are better characterized in the experimental graphs [see Figs.~\ref{fig4}(a) and (c)] due to the passive vibration isolation system used to attenuate the wave propagating to the right-hand side of the beam (see Fig. \ref{fig1}). 
Note that numerical simulations in Figs.~\ref{fig4}(b) and (d)] also show multiple reflections in the uniform part of the beam due to the fact that the algorithm didn't included any absorbing mechanism at the end of the beam. 

Now, let us discuss the sample with smaller chirp intensity, $\gamma =$ 0.03, whose expected oscillation period, according to Eq.~(\ref{eq2}), is $T_B=3.3$~ms. 
The corresponding results are shown in Fig.~\ref{fig5}.   
It is noticeable that, instead of the oscillatory behavior, the wave packet is reflected at different depths inside the beam when the central frequency $f_\mathrm{C}$ also changes within the second band. 
The wave packet penetrates deeper in the beam for increasing values of $f_\mathrm{C}$. 
This observation can be considered as far the more relevant result presented here and not only represents the mechanical analogue of the rainbow trapping effect but it also gives the location, within the mechanical structure, in which the Bloch oscillations start to develop.  
In this case the wave packet holds a ``{\em rectified}'' oscillation inside the chirped region. 
In other words, the wave traveling back firstly arrives to the interface where it is partially transmitted to the homogeneous region and partially reflected back to the chirped region, as it is seen in the left panels of Fig.~\ref{fig5}, describing the experimental characterization.
The numerical simulations of the time evolution of the wave packet, shown in the right panels in Fig.~\ref{fig5}, corroborate the experimental observations. 
One can also notice that the period observed in Figs.~\ref{fig5}(c), (e) and~(g) is increasing with the frequency up to 0.8 $T_B$; the limiting case is just the Bloch oscillation with its corresponding period $T_B$. This is a new kind of oscillation, a {\em ``rectified rainbow-Bloch oscillation''} since its period is larger with increasing frequency as the rainbow trapping effect is. 

The appearance of rainbow trapping in the beam with smaller chirping parameter can be explained in terms of the locally periodic structure using the independent rod model~\cite{GutierrezEtAl} as follows. 
In this case the system can be considered as a chirped mechanical crystal in which there is a local variation of the bandgaps along the structure. 
Therefore, the wave traveling inside this quasi-periodic crystal is gradually slowing down, as the wave frequency which is propagating is approaching the ``local" bandgaps \cite{RomeroGarciaEtAl}.
The physical mechanism of wave reflection from the point where the local index of refraction vanishes is described in Ref. \onlinecite{landau} 
Roughly, the Wannier-Stark ladders associated to a given band $j$ can be considered as a series of minigaps determined by the Bragg condition $k \ell_n=j \pi$ for cell $n$ and being $k$ the wave number; here $j=1,2,\ldots$ defines the order of the band.
For the band of interest here, $j=$1, along with Eq.~(\ref{Eq.ell}), it is possible to argue that a wave packet with central frequency $f_\mathrm{C}$ penetrates into the system up to cell $n$ given by
\begin{equation}
 n=\frac{1}{\gamma}\left(\frac{f_\mathrm{C}}{f_0} -1\right),
\label{Eq.Penetration}
\end{equation}
where $f_0=\frac{c}{2\ell_0}$ and $j=1$, corresponding to the second band in Fig.~\ref{fig2}. 
In the previous equation it was assumed that $f_\mathrm{C} \approx f_n=\frac{c}{2\ell_n}$. 
This yields the penetration of the wave packet. 
At this location it will be fully resonant and, for higher frequencies, it is completely reflected and starts oscillating with a Bloch period determined by Eq.~(\ref{eq2}).
As discussed above, the structure is too small and the effective field represented by the chirped parameter is very small holding only {\em rainbow oscillations} of the wave packet inside the chirped region of the mechanical structure.  
The penetration lengths predicted by Eq.~(\ref{Eq.Penetration}) are marked by arrows in Fig.~\ref{fig5}, showing a reasonably good agreement with the experimental observation in view of the simple model employed in its derivation.

In summary, we have reported the mechanical analogue of the electronic Bloch oscillations and the rainbow trapping effect which are predicted to appear together. 
The rainbow trapping yields the penetration length at which the wave packet will be reflected and starts oscillating with the Bloch period, when the size of the structure is large enough to hold them. 
When the structure is too small the wave packet realizes a {\em ``rectified rainbow-Bloch oscillation''} bouncing between a minigap and the interface between the structure and the uniform beam and with a period increasing with the frequency.
An analytical model and the numerical predictions based on the transfer matrix method show good agreement with experimental data.
As a potential application we can foresee the control of torsional waves propagation in metallic beams, where their penetration length can be controlled using the rainbow trapping effect. 

This work was supported by DGAPA-UNAM under project PAPIIT IN103115 and by CONACYT. 
AAL acknowledges CONACYT for the support granted to pursue his Ph. D. studies. 
G. B\'aez received CONACYT's financial support. 
RAMS received support from DGAPA-UNAM under program PASPA. 
We would like to thank M. Mart\'inez, A. Mart\'inez, V. Dom\'inguez-Rocha, E. Flores and E. Sadurn\'i for invaluable comments.  
F.C, A.C. and J.S-D. acknowledge the support by the Ministerio de Econom\'{\i}a y Competitividad of the Spanish government, and the European Union FEDER through project TEC2014-53088-C3-1-R.

\end{document}